\documentstyle[amssymb,aps,epsfig]{revtex}
\begin{document}

\title{Proton transversity and intrinsic motion of the quarks
\footnote{Talk prepared for the conference SPIN2004}}

\author{Petr Z\'{a}vada}

\address{Institute of Physics, Academy of Sciences of the Czech Republic,\\
Na Slovance 2, CZ-182 21 Prague 8\\
E-mail: zavada@fzu.cz\\
}

\maketitle

\begin{abstract}
The spin structure of the system of quasifree fermions having total angular
momentum $J=1/2$ is studied in a consistently covariant approach. Within
this model the relations between the spin functions are obtained. Their
particular cases are the sum rules Wanzura - Wilczek, Efremov - Leader -
Teryaev, Burkhardt - Cottingham and also the expression for the Wanzura -
Wilczek twist 2 term $g_{2}^{WW}$. With the use of the proton valence quark
distributions as an input, the corresponding spin functions including
transversity  are obtained.  
\end{abstract}

\section{Introduction}
In this talk some results following from the covariant quark-parton model
(QPM) will be shortly discussed, details of the model can be found in our
recent papers\cite{zav4} \cite{zav5} \cite{tra}. In this version of QPM
valence quarks are considered as quasifree fermions on mass shell. Momenta
distributions describing the quark intrinsic motion have spherical symmetry
corresponding to the constraint $J=1/2$, which represents the total angular
momentum - generally consisting of spin and orbital parts. I shall mention
the following items:

1. What sum rules follow from this approach for the spin structure functions 
$g_{1}$ and $g_{2}$?

2. How can these structure functions be obtained from the valence quark
distributions $u_{V}$ and $d_{V}$ - if the $\ SU(6)$ symmetry is assumed?

3. Why the first moment $\Gamma _{1}$ calculated in this approach can be
substantially less, than the corresponding moment calculated within the
standard, non covariant QPM, which is based on the infinite momentum frame?

4. Calculation of the transversity distribution.

\section{Model}

The model is based on the set of distribution functions $G_{k,\lambda }(%
\frac{pP}{M})$, which measure probability to find a quark in the state:%
\[
u\left( p,\lambda \mathbf{n}\right) =\frac{1}{\sqrt{N}}\left( 
\begin{array}{c}
\phi _{\lambda \mathbf{n}} \\ 
\frac{\mathbf{p}\mathbf{\sigma }}{p_{0}+m}\phi _{\lambda \mathbf{n}}%
\end{array}%
\right) ;\qquad \frac{1}{2}\mathbf{n\sigma }\phi _{\lambda \mathbf{n}%
}=\lambda \phi _{\lambda \mathbf{n}},\qquad \lambda =\pm \frac{1}{2}, 
\]%
where$\ \mathbf{n}$ coincides with the direction of the proton polarization $%
\mathbf{J}$. Correspondingly, $m$ and $p$ are quark mass and momentum,
similarly $M$ and $P$ for the proton.\textbf{\ } With the use of these
distribution functions one can define the function $H$, which in the target
rest frame reads:%
\begin{equation}
H(p_{0})=\sum_{k=1}^{3}e_{k}^{2}\Delta G_{k}(p_{0});\qquad \Delta
G_{k}(p_{0})=G_{k,+1/2}(p_{0})-G_{k,-1/2}(p_{0}),  \label{t4}
\end{equation}%
In the previous study it was shown\cite{zav4}, how the spin structure
functions can be obtained from the generic function $H$. If one assume $%
Q^{2}\gg 4M^{2}x^{2},$ then:%
\[
g_{1}(x)=\frac{1}{2}\int H(p_{0})\left( m+p_{1}+\frac{p_{1}^{2}}{p_{0}+m}%
\right) \delta \left( \frac{p_{0}+p_{1}}{M}-x\right) \frac{d^{3}p}{p_{0}}%
;\quad x=\frac{Q^{2}}{2M\nu }, 
\]%
\[
g_{2}(x)=-\frac{1}{2}\int H(p_{0})\left( p_{1}+\frac{p_{1}^{2}-p_{T}^{2}/2}{%
p_{0}+m}\right) \delta \left( \frac{p_{0}+p_{1}}{M}-x\right) \frac{d^{3}p}{%
p_{0}}. 
\]%
Let me remark, that procedure for obtaining the functions $g_{1},g_{2}$ from
the distribution $H$ is rather complex, nevertheless the task is
well-defined and unambiguous. For the transversity the corresponding
expression reads%
\[
\delta q(x)=\varkappa \int H\left( p_{0}\right) \left( Mx-\frac{p_{T}^{2}/2}{%
p_{0}+m}\right) \delta \left( \frac{p_{0}+p_{1}}{M}-x\right) \frac{d^{3}p}{%
p_{0}}, 
\]%
where the factor $\varkappa $ depends on the approach applied in the
calculation\cite{tra}.

\section{Sum rules}

One can observe, that the functions above have the same general form%
\begin{equation}
\int H(p_{0})f(p_{0},p_{1},p_{T})\delta \left( \frac{p_{0}+p_{1}}{M}%
-x\right) d^{3}p  \label{s1}
\end{equation}%
and differ only in kinematic term $\ f$. This integral, due to spheric
symmetry and presence of the $\delta -$function term, can be expressed as a
combination of the momenta: 
\begin{equation}
V_{n}(x)=\int H(p_{0})\left( \frac{p_{0}}{M}\right) ^{n}\delta \left( \frac{%
p_{0}+p_{1}}{M}-x\right) d^{3}p.  \label{s2}
\end{equation}%
One can prove\cite{zav5}, that these functions satisfy%
\[
\frac{V_{j}^{\prime }(x)}{V_{k}^{\prime }(x)}=\left( \frac{x}{2}+\frac{%
x_{0}^{2}}{2x}\right) ^{j-k};\qquad x_{0}=\frac{m}{M}. 
\]%
This relation then gives possibility to obtain integral relations between
different functions having form (\ref{s2}) or (\ref{s1}), in particular for $%
g_{1}(x)$ and $g_{2}(x)$ one gets for $m\rightarrow 0$:

\[
g_{2}(x)=-g_{1}(x)+\int_{x}^{1}\frac{g_{1}(y)}{y}dy, 
\]%
\[
g_{1}(x)=-g_{2}(x)-\frac{1}{x}\int_{x}^{1}g_{2}(y)dy. 
\]%
Another rule, which is obtained in this approach, reads:%
\[
\int_{0}^{1}x^{\alpha }\left[ \frac{\alpha }{\alpha +1}g_{1}(x)+g_{2}(x)%
\right] dx=0, 
\]%
which is valid for \textit{any} $\alpha $, for which the integral exists.
For $\alpha =2,4,6,...$ the relation corresponds to the Wanzura - Wilczek
sum rules\cite{wawi}. Other special cases correspond to the Burkhardt -
Cottingham\cite{buco} ($\alpha =0$) and the Efremov - Leader - Teryaev\cite%
{elt} (ELT, $\alpha =1$) sum rules. For the transversity one gets%
\[
\delta q(x)=2\varkappa \left( g_{1}(x)+\int_{x}^{1}\frac{g_{1}(y)}{y}%
dy\right) . 
\]

\section{Valence quarks}

Now I shall apply the suggested approach to the description of the real
proton. For simplicity I assume:

1) Spin contribution from the sea of quark-antiquark pairs and gluons can be
neglected, so the proton spin is generated only by the valence quarks. The
negligible contribution from the quark sea was recently reported in\cite{her}%
.

2) In accordance with the non-relativistic \textit{SU(6)} approach, the spin
contribution of individual valence terms is given by the fractions:%
\[
s_{u}=4/3,\qquad s_{d}=-1/3. 
\]%
If the symbols $h_{u}$ and $h_{d}$ denote momenta distributions of the
valence quarks in the proton rest frame, which are normalized as%
\[
\frac{1}{2}\int h_{u}(p_{0})d^{3}p=\int h_{d}(p_{0})d^{3}p=1, 
\]%
then the generic distribution (\ref{t4}) reads%
\begin{equation}
H(p_{0})=\sum e_{j}^{2}\Delta h_{j}(p_{0})=\left( \frac{2}{3}\right) ^{2}%
\frac{2}{3}h_{u}(p_{0})-\left( \frac{1}{3}\right) ^{2}\frac{1}{3}%
h_{d}(p_{0}).  \label{t71}
\end{equation}

\begin{figure}
\begin{center}
\epsfig{file=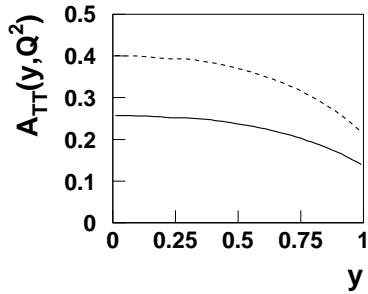, height=5cm}
\end{center}
\caption{Double spin asymmetry at $Q^{2}=4GeV/c$ is calculated using two
transversity approaches: Interference effects are attributed to quark level
only \textit{(solid line)}. Interference effects at parton-hadron transition
stage are included in addition \textit{(dashed line)}, this curve represents
upper bound only. }
\label{fgr1}
\end{figure}

In the paper\cite{zav1}, using a similar approach, I studied also the
unpolarized structure functions. Structure function $F_{2}$ can be expressed
as%
\[
F_{2}(x)=x^{2}\int G(p_{0})\frac{M}{p_{0}}\delta \left( \frac{p_{0}+p_{1}}{M}%
-x\right) d^{3}p;\qquad G(p_{0})=\sum_{q}e_{q}^{2}h_{q}(p_{0}). 
\]%
On the other hand, for proton valence quarks one can write%
\[
F_{2}(x)=\frac{4}{9}xu_{V}(x)+\frac{1}{9}xd_{V}(x), 
\]%
so combination of the last two relations gives:%
\[
q_{V}(x)=x\int h_{q}(p_{0})\frac{M}{p_{0}}\delta \left( \frac{p_{0}+p_{1}}{M}%
-x\right) d^{3}p;\qquad q=u,d. 
\]%
Since this is again the integral having the structure (\ref{s1}), one can
apply the technique of integral transforms and (instead of relations between 
$g_{1},g_{2},\delta q$) obtain the relations between $g_{j}^{q},\delta q$
and $q_{V}$. For $m\rightarrow 0$ these relations read: 
\[
g_{1}^{q}(x)=\frac{\cos \eta _{q}}{2}\left( \allowbreak
q_{V}(x)-2x^{2}\int_{x}^{1}\frac{q_{V}(y)}{y^{3}}dy\right) , 
\]%
\[
g_{2}^{q}(x)=\frac{\cos \eta _{q}}{2}\left( -\allowbreak \allowbreak
q_{V}(x)+3x^{2}\int_{x}^{1}\frac{q_{V}(y)}{y^{3}}dy\right) , 
\]%
\[
\delta q(x)=\varkappa \cos \eta _{q}\left( q_{V}(x)-x^{2}\int_{x}^{1}\frac{%
q_{V}(y)}{y^{3}}dy\right) , 
\]%
where $\cos \eta _{q}$ is corresponding $SU(6)$ factor. Now, taking quark
charges and the $SU(6)$ factors as in Eq. (\ref{t71}), one can directly
calculate $g_{1},g_{2}$ and $\delta q$\ only using the input on the valence
quark distributions $q_{V}=u_{V},d_{V}$. More detailed discussion of
obtained results is done in our cited papers. Here I mention our (A.Efremov,
O.Teryaev and P.Zavada) quite recent result on double spin asymmetry $%
A_{TT}(y,Q^{2})$, which is shown in Fig. (\ref{fgr1}).

Calculation is done in an accordance with the recent study\cite{pax}, but
with the use of quark trasversities obtained in the presented approach,
which gives slightly lower estimation of the $A_{TT}$. It is expected, that
this function could be accessible in the PAX experiment\cite{paxc}, which
will measure spin asymmetry of the lepton pairs produced in collisions of
transversely polarized\ protons and antiprotons.

Further, in this talk I want concentrate on the discussion and explanation,
why intrinsic quark motion substantially reduces the first moment of the
spin function $g_{1}$. I have shown\cite{zav4}, that

\begin{equation}
\Gamma _{1}\equiv \int g_{1}(x)dx=\frac{1}{2}\int H(p_{0})\left( \frac{1}{3}+%
\frac{2m}{3p_{0}}\right) d^{3}p,  \label{v1}
\end{equation}%
which, in the $SU(6)$ approach gives%
\[
\frac{5}{18}\geq \Gamma _{1}\geq \frac{5}{54},
\]%
where left limit is valid for the static and right one for massless quarks.
In other words, it seems: 
\[
more\ intrinsic\ motion\Leftrightarrow less\ spin
\]%
All right, this is a mathematical result, but how to understand it from the
point of view of physics?

First, forget structure functions for a while and calculate completely
another task.\ Let me remind general rules concerning angular momentum in
quantum mechanics:

1) Angular momentum consist of orbital and spin part: \textbf{j=l+s}

2) In the relativistic case \textbf{l} and \textbf{s} are not conserved
separately, only total angular momentum \textbf{j} is conserved. So, one can
have pure states of $j(j^{2},j_{z})$ only, which are for fermions with $%
s=1/2 $ represented by the relativistic spheric waves\cite{lali}:

\[
\psi _{jlj_{z}}\left( \mathbf{p}\right) =\frac{1}{\sqrt{2p_{0}}}\left( 
\begin{array}{c}
i^{-l}\sqrt{p_{0}+m}\Omega _{jlj_{z}}\left( \frac{\mathbf{p}}{p}\right) \\ 
i^{-l^{\prime }}\sqrt{p_{0}-m}\Omega _{jl^{\prime }j_{z}}\left( \frac{%
\mathbf{p}}{p}\right)%
\end{array}%
\right) ;\qquad j=l\pm \frac{1}{2},\qquad l^{\prime }=2j-l, 
\]%
\begin{eqnarray*}
\Omega _{l+1/2,l,j_{z}}\left( \frac{\mathbf{p}}{p}\right) &=&\left( 
\begin{array}{c}
\sqrt{\frac{j+j_{z}}{2j}}Y_{l,j_{z}-1/2} \\ 
\sqrt{\frac{j-j_{z}}{2j}}Y_{l,j_{z}+1/2}%
\end{array}%
\right) ,\qquad \\
\Omega _{l^{\prime }-1/2,l^{\prime },j_{z}}\left( \frac{\mathbf{p}}{p}%
\right) &=&\left( 
\begin{array}{c}
-\sqrt{\frac{j-j_{z}+1}{2j+2}}Y_{l^{\prime },j_{z}-1/2} \\ 
\sqrt{\frac{j+j_{z}+1}{2j+2}}Y_{l^{\prime },j_{z}+1/2}%
\end{array}%
\right) .
\end{eqnarray*}%
This wavefunction is simplified for the state with total angular momentum
(spin) equal 1/2:

\[
j=j_{z}=\frac{1}{2},\qquad l=0\qquad \Rightarrow \qquad l^{\prime }=1, 
\]%
\[
Y_{00}=\frac{1}{\sqrt{4\pi }},\qquad Y_{10}=i\sqrt{\frac{3}{4\pi }}\cos
\theta ,\qquad Y_{11}=-i\sqrt{\frac{3}{8\pi }}\sin \theta \exp \left(
i\varphi \right) , 
\]%
which gives%
\[
\psi _{jlm}\left( \mathbf{p}\right) =\frac{1}{\sqrt{8\pi p_{0}}}\left( 
\begin{array}{c}
\sqrt{p_{0}+m}\left( 
\begin{array}{c}
1 \\ 
0%
\end{array}%
\right) \\ 
-\sqrt{p_{0}-m}\left( 
\begin{array}{c}
\cos \theta \\ 
\sin \theta \exp \left( i\varphi \right)%
\end{array}%
\right)%
\end{array}%
\right) . 
\]%
Let me remark, that $j=1/2$ is minimum angular momentum for particle with $%
s=1/2.$ Now, one can easily calculate the average contribution of the spin
operator to the total angular momentum:

\[
\Sigma _{3}=\frac{1}{2}\left( 
\begin{array}{cc}
\sigma _{3} & \cdot \\ 
\cdot & \sigma _{3}%
\end{array}%
\right) \Rightarrow 
\]%
\[
\psi _{jlm}^{\dagger }\left( \mathbf{p}\right) \Sigma _{3}\psi _{jlm}\left( 
\mathbf{p}\right) =\frac{1}{16\pi p_{0}}\left[ \left( p_{0}+m\right) +\left(
p_{0}-m\right) \left( \cos ^{2}\theta -\sin ^{2}\theta \right) \right] 
\]%
If $a_{p}$ is the probability amplitude of the state $\psi _{jlm}$, then 
\begin{equation}
\left\langle \Sigma _{3}\right\rangle =\int a_{p}^{\star }a_{p}\psi
_{jlm}^{\dagger }\left( \mathbf{p}\right) \Sigma _{3}\psi _{jlm}\left( 
\mathbf{p}\right) d^{3}p=\frac{1}{2}\int a_{p}^{\star }a_{p}\left( \frac{1}{3%
}+\frac{2m}{3p_{0}}\right) p^{2}dp,  \label{v2}
\end{equation}%
which means, that:

\textit{i)} For the fermion at rest ($p_{0}=m$) we have $j=s=1/2,$ which is
quite comprehensible, since without kinetic energy no orbital momentum can
be generated.

\textit{ii)}\ For the state in which $p_{0}\geq m$, we have in general: 
\[
\frac{1}{3}\leq \frac{\left\langle s\right\rangle }{j}\leq 1. 
\]%
where left limit is valid for the energetic (or massless) fermion, $p_{0}\gg
m$. In other words, in the states $\psi _{jlm}$ with $p_{0}>m$ part of the
total angular momentum $j=1/2$ is \textit{necessarily }created by orbital
momentum. This is a simple consequence of quantum mechanics.

Now, one can compare integrals (\ref{v1}) and (\ref{v2}). Since both
integrals involve the same kinematic term, the interpretation of dependence
on ratio $m/p_{0}$ in (\ref{v2}) is valid also for (\ref{v1}). Otherwise,
the comparison is a rigorous illustration of the statement, that $\Gamma
_{1} $ measures contributions from quark spins (and not their total angular
momenta).

\section{Summary}

I have studied spin functions in the system of quasifree fermions having
total spin $J=1/2$ - representing a covariant version of naive QPM. The main
results are:

1) Spin functions $g_{1}$ and $g_{2}$ depend on intrinsic motion. In
particular, the momenta $\Gamma _{1}$ corresponding to the static (massive)
fermions and massless fermions, can differ significantly: $\Gamma _{1}(m\ll
p_{0})/\Gamma _{1}(p_{0}\approx m)=1/3$. It is due to splitting of angular
momentum into spin and orbital part, as soon as intrinsic motion is present.

2) $g_{1}$ and $g_{2}$ are connected by a simple transformation, which is
for $m\rightarrow 0$ identical to Wanzura - Wilczek relation for twist-2
term of the $g_{2}$ approximation. Relations for the $n-th$ momenta of the
structure functions have been obtained, their particular cases are identical
to known sum rules: Wanzura - Wilczek ($n=2,4,6...$), Efremov - Leader -
Teryaev ($n=1$) and Burkhardt - Cottingham ($n=0$).

3) Model has been applied to the proton spin structure, assuming proton spin
is generated only by spins and orbital momenta of the valence quarks with 
\textit{SU(6)} symmetry and for quark effective mass $m\rightarrow 0$. Using
a\ known parameterization of the valence terms as the input, the functions $%
g_{1},$ $g_{2}$ and the transversity $\delta q$ are obtained.

\end{document}